\pdfoutput=1

\documentclass[aps,prl,preprint, superscriptaddress, showpacs]{revtex4-1}

\usepackage{graphicx, color}
\usepackage{dcolumn}
\usepackage{bm}
\usepackage{amsmath}
\usepackage{amssymb}
\usepackage{verbatim}
\usepackage{appendix}
\usepackage{natbib}
\usepackage{caption}
\usepackage{subcaption}
\usepackage{wrapfig}
\usepackage[font=small,labelfont=bf]{caption}
\usepackage{multirow}

\begin{document}


\title{Decoherence-protected transitions of nitrogen vacancy centers in 99\% $^{13}$C diamond}



\author{Anna J. Parker}
\affiliation{Department of Chemistry, University of California, Berkeley, California 94720}
\affiliation{Materials Science Division, Lawrence Berkeley National Laboratory, USA}
\author{Haijing Wang}
\affiliation{Department of Chemistry, University of California, Berkeley, California 94720}
\affiliation{Materials Science Division, Lawrence Berkeley National Laboratory, USA}
\author{Yiran Li}
\affiliation{Department of Chemistry, University of California, Berkeley, California 94720}
\author{Alexander Pines}
\affiliation{Department of Chemistry, University of California, Berkeley, California 94720}
\affiliation{Materials Science Division, Lawrence Berkeley National Laboratory, USA}
\author{Jonathan P. King}
\email{jpking@berkeley.edu}
\affiliation{Department of Chemistry, University of California, Berkeley, California 94720}
\affiliation{Materials Science Division, Lawrence Berkeley National Laboratory, USA}



\date{\today}

\begin{abstract}
Nitrogen vacancy (NV$^{-}$) color centers in diamond are a prime candidate for use in quantum information devices, owing to their spin-1 ground state, straightforward optical initialization and readout, and long intrinsic coherence times in a room-temperature solid. While the $^{13}$C nuclear spin is often a dominant source of magnetic noise, we observe transitions between electron-nuclear hyperfine states of NV$^-$ centers in 99\% $^{13}$C diamond that are robust to decoherence. At magnetic field strengths ranging from 550 - 900 G, these transitions are observable by optically detected magnetic resonance (ODMR), and exhibit linewidths narrowed by factors as high as $\sim$130 at room temperature over typical electron-type transitions observed from this spin system. We anticipate the use of these decoherence-protected transitions, in combination with dynamical decoupling methods, for storage of quantum information.

\end{abstract}


\maketitle
\section{Introduction}

The nitrogen vacancy (NV$^-$) center is often regarded as a trapped ion in diamond, with localized electronic states resulting from its insulating environment and an optical transition distinct from the diamond bandgap. Its ground state has a total electronic spin of 1, and the spin-spin interaction within the defect results in an energy splitting between the $m_S$ = 0 and $m_S$ = $\pm$ 1 spin states. Due to optical dynamics and selection rules of NV$^-$ centers, the $m_S$ = 0 spin state fluoresces more intensely than the $m_S$ = $\pm$ 1 spin states and also becomes nearly completely populated upon optical illumination. As a result, the m$_S$ = 0 spin state is referred to as ``bright" and the m$_S$ = $\pm$1 spin states are referred to as ``dark". This spin-dependent fluorescence forms the basis for the method of optically-detected magnetic resonance (ODMR) of NV$^-$ centers. These properties, in combination with the details of the energy level structure of the NV$^-$ center, allow for its use in optical sensing of temperature, pressure, magnetic and electric fields \cite{kucsko13,doherty2014,doherty14,maze08,degen08,dolde11}; for atomic-scale nuclear magnetic resonance (NMR) and magnetic field imaging, as well as for quantum information processing (QIP)\cite{mamin13,staud13,gopi08,sushkov14,epstein05,nemoto14}. The NV$^-$ center is competitive as a quantum bit (qubit) candidate due to the ease of manipulating its optical and spin degrees of freedom as well as the favorable properties and versatile fabrication of the diamond. One significant challenge in the implementation of a working quantum computer common to many implementations is producing a qubit that is sufficiently decoupled from its environment but may also be interrogated efficiently \cite{divincenzo00}. In this work we address the issue of spin decoherence of NV$^-$ centers.


The sources of decoherence and their interactions with the qubit have been studied for many qubit candidates, including photons, phosphorous and silicon nuclear spins in silicon, quantum dots, trapped atoms and ions, superconducting circuits, and the NV$^-$ center in diamond \cite{ladd10,morton11}. Methods of extending coherence times generally include physically removing the sources of decoherence, for example by isotopic enrichment of the host material \cite{bala09,itoh14}, and dynamical decoupling (DD) pulse sequences inspired by the Hahn echo that remove interference from static and fluctuating sources of noise \cite{dobrov13}. Another approach to achieving useful coherence times is to use transitions that are inherently protected from sources of decoherence. One form of these transitions is generally termed Zero First-Order Zeeman (ZEFOZ) for satisfying the condition $\frac{\partial\nu}{\partial B} \rightarrow 0$, in other words when the first-order magnetic field ($B$) dependence of a transition with frequency $\nu$ goes to zero. These are typically found in systems that possess internal interactions (i.e. hyperfine interactions or zero field splittings) that do not commute with the Zeeman interaction. For spin qubits with such internal structure, magnetic field conditions exist where the Zeeman energy cancels the hyperfine or zero-field energy, resulting in an avoided crossing or level anti-crossing (LAC). Approaching the regime of the LAC, the magnetic field dependence of spin states becomes nonlinear, resulting in ZEFOZ conditions where the transition frequency is desensitized to perturbations in the magnetic field. Such phenomena may also be understood by realizing that the first-order Zeeman shift of a transition frequency is its effective gyromagnetic ratio, $\gamma_{eff}$. Spin transitions with weak $\gamma_{eff}$ may be characterized by long coherence times because they couple weakly to local magnetic field fluctuations that result in dephasing. This is the same reasoning used to explain why nuclear spin coherence times generally exceed electron spin coherence times by orders of magnitude: their gyromagnetic ratios are weaker than that of the electron by a factor of 10$^3$-10$^4$. 

The use of ZEFOZ transitions to extend coherence was originally introduced by Longdell and coworkers with rare-earth metal ion doped materials \cite{fraval04}. Extended coherence times were demonstrated for Pr$^{3+}$:Y$_2$SiO$_5$ and  Pr$^{3+}$:La$_2$(WO$_4$)$_3$ for enhancements of $\sim$630 using the ZEFOZ method alone \cite{fraval04,lovric11}. Building upon these gains has led to some of the longest quantum memory times currently known \cite{zhong15,heize13}. Furthermore, the ZEFOZ method is broadly applicable. In trapped $^9$Be$^+$ ions, the ZEFOZ transition between hyperfine levels $\mid F = 2, m_F = 0 \rangle \leftrightarrow \mid F = 1, m_F = 1 \rangle$ has a coherence time of $\sim$10 s, which is an improvement of 5 orders of magnitude over transitions in the same system whose magnetic field dependence is linear \cite{langer05}. Use of a ZEFOZ transition in bismuth-doped natural silicon results in over two orders of magnitude increase in coherence time to $\sim$90 ms \cite{wolf13}. Such transitions have already been applied to extend coherence of NV$^-$ centers in 1.1\% $^{13}$C diamond: Lesik and coworkers reported an 8-fold increase in coherence time for bulk NV$^-$ centers by tuning to a ZEFOZ point at zero magnetic field induced by the hyperfine interaction with $^{14}$N of the defect \cite{lesik15}. This gain in coherence was shown to diminish to a factor of 2 for shallow implanted NV$^-$ centers or those in nanodiamonds where the defect is subject to stronger electric field fluctuations. Lastly, Xu et al. employed continuous-wave DD of dressed NV$^-$ center spin states, which essentially utilized ZEFOZ transitions in the rotating frame, in 1.1\% $^{13}$C diamond to achieve enhancements of $\sim$20 in spin coherence time \cite{xu12}. 

In the current study we apply the principles of the ZEFOZ technique to the mixed states of NV$^{-}$ centers in $^{13}$C-enriched diamond and observe a maximum reduction of linewidth by a factor of 133.7 from pure electron-type spin transitions typical of the system. We are able to explain much of the spectral behavior of this spin system observed by continuous-wave (CW)-ODMR using a simple model of four spins: the NV$^-$ and its three nearest-neighbor $^{13}$C nuclei. Due to our method of detection, we cannot observe transitions directly at ZEFOZ points found within this spin system, meaning that while coherence is enhanced significantly, the transitions are protected from decoherence (``deocherence-protected transitions" or DPTs) rather than true ZEFOZ transitions.
  
\section{Decoherence of NV$^-$  centers in a nuclear spin bath}
The spin dynamics of NV$^{-}$ centers are highly sample dependent, determined largely by the spin content of the diamond host as well as by the morphology and treatment of the material. Tuning these conditions may result in coherence times as short as hundreds of nanoseconds and as long as milliseconds \cite{acosta09,bala09,jelezko04,trusheim14,mizu09,doherty13}. Consequently, extensive effort has been invested in the development of diamond samples ideal for applications that require long spin coherence time. 

\begin{figure}[h]
	 \includegraphics[width=0.6\textwidth]{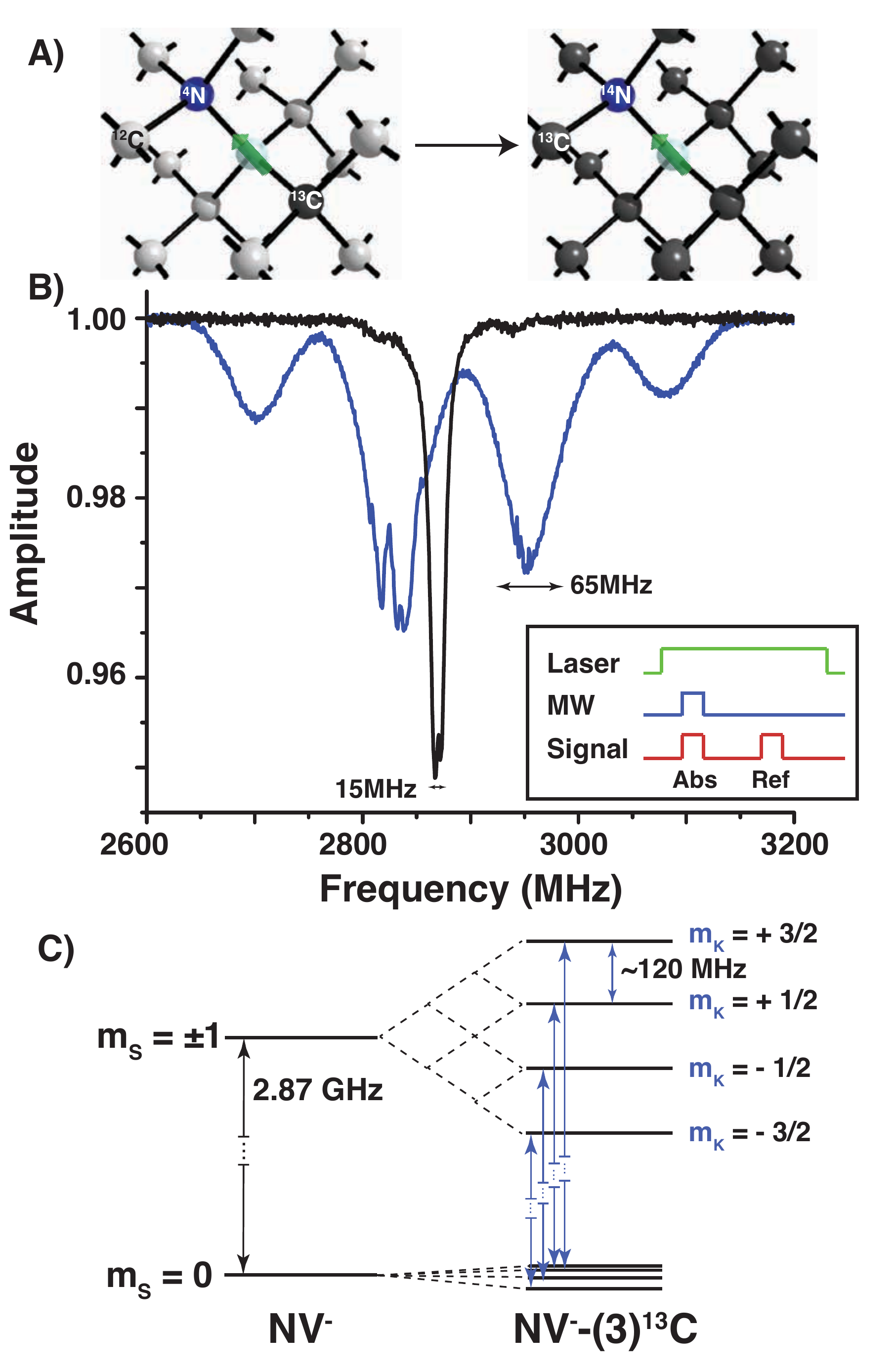}
	\caption{The NV-(3)$^{13}$C electron-nuclear spin system. (A) Schematic of the NV$^{-}$ center in 1.1\% and 99\% $^{13}$C diamond. $^{14}$N, $^{12}$C, $^{13}$C, and the lattice vacancy are depicted as blue, gray, dark gray, and transparent cyan spheres, respectively. In fully-enriched $^{13}$C diamond, all lattice sites around the NV$^-$ center are occupied by $^{13}$C nuclei. (B) EarthÕs field (~0.5G) CW-ODMR spectra of ensembles of NV$^{-}$ centers in 1.1\% (black) and 99\% (blue) $^{13}$C diamond. The inset describes the experiment used to acquire all spectra presented in this study (See Experimental). (C) Energy level diagram of the NV$^{-}$ center in the absence of \textsuperscript{13}C nuclei in the first shell (black) and with three \textsuperscript{13}C nuclei in the first shell (blue). The spin quantum number of the NV$^-$ is denoted by m$_S$, whereas the spin quantum number of the total nuclear spin of the three first-shell $^{13}$C nuclei is denoted by m$_K$. }
	\label{fig:nvc13}
\end{figure}

In bulk single crystalline diamond, the main sources of decoherence for NV$^{-}$ centers may be $^{13}$C nuclear spins, P1 centers (a spin-$^1/_2$ electronic defect arising from a substitutional nitrogen atom in the diamond lattice), or neutral NV$^{0}$ centers \cite{doherty13}. Because our sample has low nitrogen content and high $^{13}$C content, we consider only the effect of the nuclear spin bath on NV$^{-}$ linewidth ($\Delta\nu$). We use this linewidth as an indicator of coherence time ($T_2$) by the relation $T_2^* \sim (\Delta\nu)^{-1}$, where $T_2^*$, the inhomogeneous dephasing time, is less than $T_2$. Mizuochi et al. reported  NV$^{-}$ resonances as narrow as 18 kHz by isotopic purification to 99.9997\% $^{12}$C diamond. In comparison, a dense nuclear spin bath such as that in 99\% $^{13}$C diamond can lead to linewidths ranging from 50-70 MHz (shown in blue, Figure \ref{fig:nvc13}B).This difference is the consequence of broadening by a range of hyperfine interactions from $^{13}$C nuclei occupying all sites surrounding the NV$^-$ defect in the diamond lattice.

We explain some of the behavior exhibited by an ensemble of NV$^-$  centers in $^{13}$C-enriched diamond (Figure \ref{fig:nvc13}A) in this paper. The ensemble is approximated as a four-spin ``molecule" composed of the NV$^-$  center and the three nearest-neighbor nuclei, called ``first-shell nuclei". The operators $S_z$, $S_x$, and $S_y$ describe the electron spin angular momentum, individual nuclear spin angular momenta are described by the operators $I_{nx}$, $I_{ny}$, and $I_{nz}$, and the total nuclear spin operators are given by $K_x$, $K_y$, and $K_z$. With the applied field along the NV$^-$ axis, the ground state spin Hamiltonian ($H_{gs}$) includes the electron spin Zeeman term ($\gamma_{NV} B_zS_z$), the ground state zero-field splitting interaction ($D_{gs}$ = 2.87 GHz), and the ground state hyperfine Hamiltonian ($H_{HF,gs}$) of the NV$^-$ center interacting with three first-shell spin-${\frac{1}{2}}$ nuclei,  
\begin{equation*}
H_{gs} = \gamma_{NV} B_zS_z+D_{gs} \left(S_z^2-\frac{1}{3}S^2\right)+H_{HF,gs}
\end{equation*}
where 
\begin{equation*}
H_{HF,gs}= (S\cdot A\cdot I_1)+R_z^{(i\phi)}(S\cdot A\cdot I_2)R_z^{(-i\phi)}+R_z^{(i2\phi)}(S\cdot A\cdot I_3)R_z^{(-i2\phi)}
\end{equation*}.
The gyromagnetic ratio of the NV$^-$ ($\gamma_{NV}$) is $\sim$2.8 MHz/G. We construct the Hamiltonian in the frame of the defect axis because our experiment only detects transitions in the NV$^{-}$ spin state along that axis. With the NV$^-$-$^{13}$C hyperfine tensor ($A$) projected onto the NV$^-$ axis, taking the nucleus $^{13}$C$_1$ to lie in the $xz$ plane, the hyperfine term for one nuclear spin in $H_{HF,gs}$ becomes 
\begin{equation*}
(S\cdot A\cdot I_1) = A_{xx}S_xI_{1x} + A_{yy}S_yI_{1y} + A_{zz}S_zI_{1z} + A_{xz}(S_zI_{1x} + S_xI_{1z})
\end{equation*}
with hyperfine tensor elements \cite{shim13}: A$_{xx}$ = 166.9 MHz, A$_{yy}$ = 122.9 MHz, A$_{zz}$ = 90 MHz, and A$_{xz}$ = -90 MHz. We obtain the hyperfine interactions with the remaining two nuclei by incremental rotations of $\phi$ = ${\frac{2\pi}{3}}$ about the NV$^-$ axis ($R_z^{(i\phi)}$).

The magnetic field dependence of the eigenvalues of the four-spin model is shown in Figure \ref{fig:elevels}. This system has a total of 24 eigenstates where each electron spin state is split into eight electron-nuclear spin states. We will refer to the sets of eigenstates where the z projection of the NV$^-$ spin angular momentum is approximately 0, -1, or 1 as the m$_S$ = 0, -1, or +1 ``manifolds". In the limit of B = 0, the three hyperfine interactions from the first-shell spin-$\frac{1}{2}$ nuclei split the $m_S$ = $\pm$1 manifolds each into quartet structures. When combining three nuclei of spin-$\frac{1}{2}$, there is only one combination that gives a total nuclear spin of $\pm\frac{3}{2}$, but three possible combinations each that give a total nuclear spin of $\pm\frac{1}{2}$. This gives rise to the 1:3:3:1 intensity ratio of resonances observed in the zero field spectrum, shown in Figure \ref{fig:nvc13}B. The transitions leading to this quartet structure (given in blue in the energy-level diagram in Figure \ref{fig:nvc13}C) are inhomogeneously broadened by the various configurations of weak hyperfine interactions of nuclei occupying sites at a distance of ~3-8 angstroms \cite{smeltz11,gali08} from the defect. In contrast, the zero field spectrum of an ensemble of NV$^-$ centers in 1\% $^{13}$C diamond (shown in black, Figure \ref{fig:nvc13}B) has one primary resonance, the zero field resonance (ZFR), split by crystal strain of approximately 3 MHz. The primary resonance is accompanied by two broadened lines of weakened intensity at $D_{gs} - 56.9$ MHz and $D_{gs} + 70.7$ MHz, corresponding to the small probability of finding a NV$^{-}$ center coupled to a first-shell $^{13}$C nucleus \cite{simanov13}. It should be noted that the linewidth of this particular ensemble of NV$^{-}$ centers in 1\%$^{13}$C diamond is large for NV$^{-}$ centers in diamond of this composition at $\sim$15 MHz. This is attributed to the proximity of the defects to the surface (no greater than 100 nm) as surface strain and spin defects are known to produce electric field gradients and random fluctuating magnetic fields that lead to rapid dephasing.  

The eigenstates of the m$_S$ = 0 manifold are also split into four two-fold degenerate levels (see Figure \ref{fig:elevels}B) in the limit of B = 0 due to the transverse terms of the hyperfine tensor. As a result the structure of the $m_S$ = 0 manifold cannot be described by the z-projection of total nuclear spin angular momentum. Even at magnetic fields far from any LAC regime, these eigenstates exhibit nonlinear magnetic field dependence and sufficient mixing for the observation of narrowed DPTs by CW-ODMR spectroscopy. 

\section{Results and Discussion}
In CW-ODMR of NV$^{-}$ centers, a transition in NV$^{-}$ spin state is saturated and observed as a decrease in fluorescence intensity. The contrast, or normalized difference in fluorescence intensity, of these transitions may be explained as a depopulation of the m$_S$ = 0 spin state \cite{wang14}. Since mixed eigenstates are not characterized by the pure Zeeman basis, but linear combinations of these states, the notion of ``bright" and ``dark" NV$^{-}$ spin states must therefore be redefined as the spin states with the most or least m$_S$ = 0 spin state character. Due to small mixing of the electronic spin state in the hyperfine states of the $m_S$ = 0 manifold, transitions that are nominally forbidden become weakly allowed and have sufficient contrast to be detected by ODMR. 

At magnetic field strengths where the influence of electron-nuclear hyperfine interactions and the zero-field splitting remains significant, all transitions in the four-spin system are possible. In order to gain insight into the nature of the decoherence protected transitions observed, we reduce the number of possible transitions to consider by characterizing them with an intensity factor $\kappa$, which evaluates the probability of their observation by CW-ODMR. $\kappa$ is determined by the product of the transition matrix element (TME), a term describing the difference in population of final and initial states (${\Delta\langle \rho \rangle}$), and a term describing the optical contrast (${\Delta}$${\langle S_z^2 \rangle}$) of the transition:
\begin{equation*}
\kappa = (TME)(\Delta\langle \rho \rangle)(\Delta\langle S_z^2 \rangle)
\end{equation*}
Where 
 \begin{equation*}
TME = \langle \varphi_f | (\gamma_{NV}(S_x + S_y + S_z)+\gamma_I(K_x + K_y + K_z)) | \varphi_i \rangle
\end{equation*}
 \begin{equation*}
\Delta\langle \rho \rangle = \langle \varphi_f | \rho | \varphi_f \rangle - \langle \varphi_i | \rho | \varphi_i \rangle
\end{equation*}
 \begin{equation*}
\Delta\langle S_z^2 \rangle = \langle \varphi_f | S_z^2 | \varphi_f \rangle - \langle \varphi_i | S_z^2 | \varphi_i \rangle
\end{equation*}
Here, because optical contrast in ODMR is determined by a change in electron spin state, we define the optical contrast term as the difference in expectation value of S${_z^2}$ in the final ($\varphi_f $) and initial ($\varphi_f $) eigenstates. The transition matrix element is calculated as the inner product of the final and initial eigenstates of the spin angular momentum operators with the electron ($S_x, S_y, S_z$) and total nuclear ($K_x, K_y, K_z$) spins, weighted by their gyromagnetic ratios. We consider all projections of angular momentum because the orientation of our microwave excitation is not exactly known. In this way we do not unintentionally exclude transitions that could explain the behavior in the ODMR spectrum.  Finally, ${\Delta\langle \rho \rangle}$ is the difference in expectation value of the density matrix ($\rho$) of the eigenstates involved in the transition. To describe the optical pumping, we construct a density matrix where the eight $m_S$ = 0 sublevels are equally populated and the nuclear spin states are thus unpolarized ($\rho = (E - S_z^2)_{NV}\otimes E_{^{13}C_1}\otimes E_{^{13}C_2}\otimes E_{^{13}C_3}$). 

\begin{figure}[h]
	 \includegraphics[width=\textwidth]{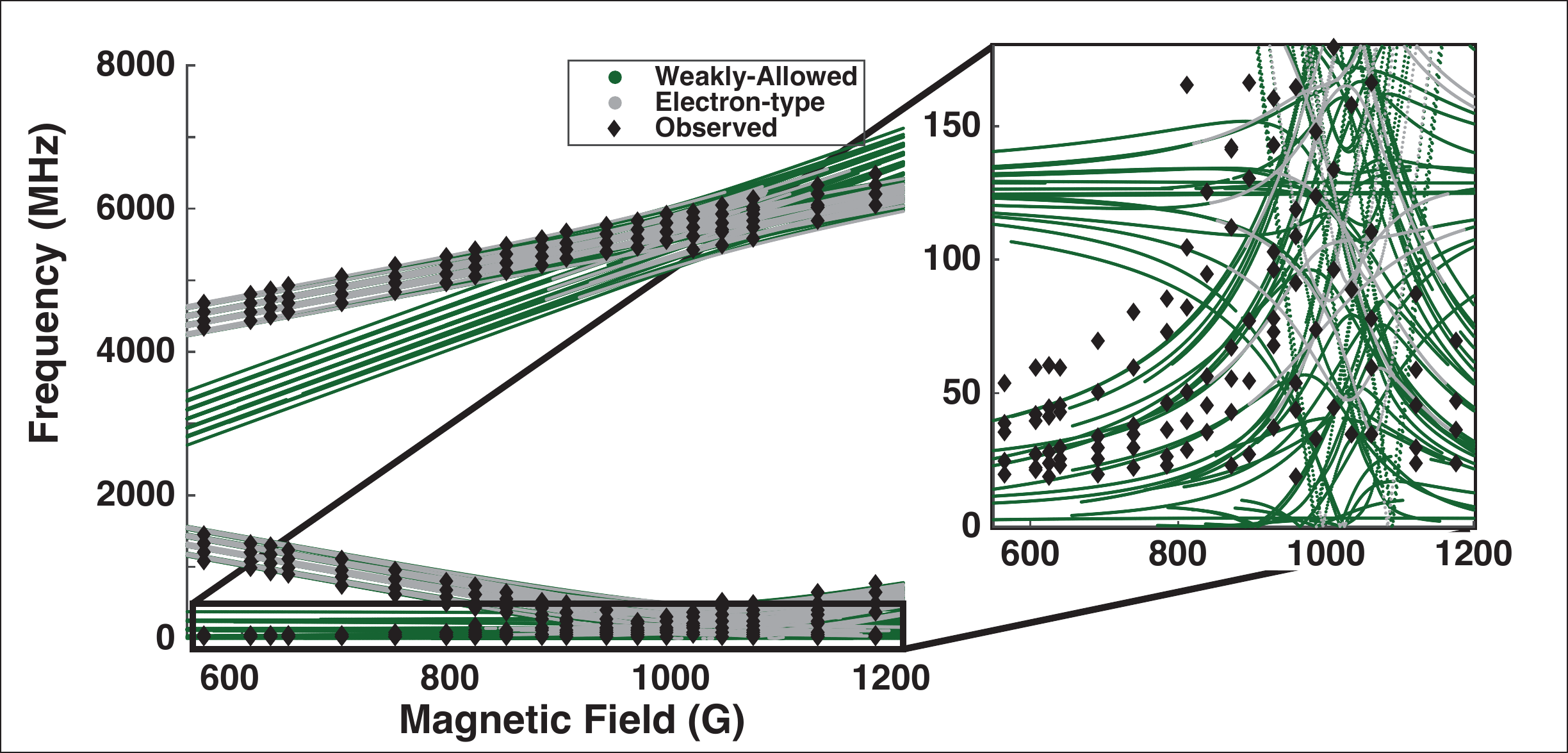}
	\caption{Predicted and observed transitions of the NV$^-$-(3)$^{13}$C spin system. Predicted transitions have intensity factors ($\kappa$) greater than 10$^{-6}$. This is 67.5 \% of all possible transitions. The frequency range of the DPTs is magnified at the right for a more detailed view. Many weakly-allowed transitions with first-order magnetic field dependence terms (i.e. $\frac{\partial\nu}{\partial B}$, or $\gamma_{eff}$ orders of magnitude lower than $\gamma_{NV}$ are predicted in this magnetic field regime.}
	\label{fig:predtrans}
\end{figure}

In Figure \ref{fig:predtrans}, the transitions observed by CW-ODMR spectroscopy are summarized with transitions we predict using $\kappa$ greater than 10$^{-6}$, which was determined empirically to best fit the data. A more detailed view of the DPTs is given in the inset to the right. Roughly two types of transitions may be distinguished: high intensity m$_S$ = 0 to m$_S$ = $\pm$1 electron-type transitions (shown in gray), and low intensity DPTs (shown in green). As discussed, transitions of the latter group have much narrower linewidths than the electron-type transitions between NV$^{-}$ spin states due to weakened effective gyromagnetic ratios of the eigenstates near LACs, indicating longer dephasing (T$_2$*) and potentially coherence (T$_2$) times. 



\begin{figure}[h]
	 \includegraphics[width=\textwidth]{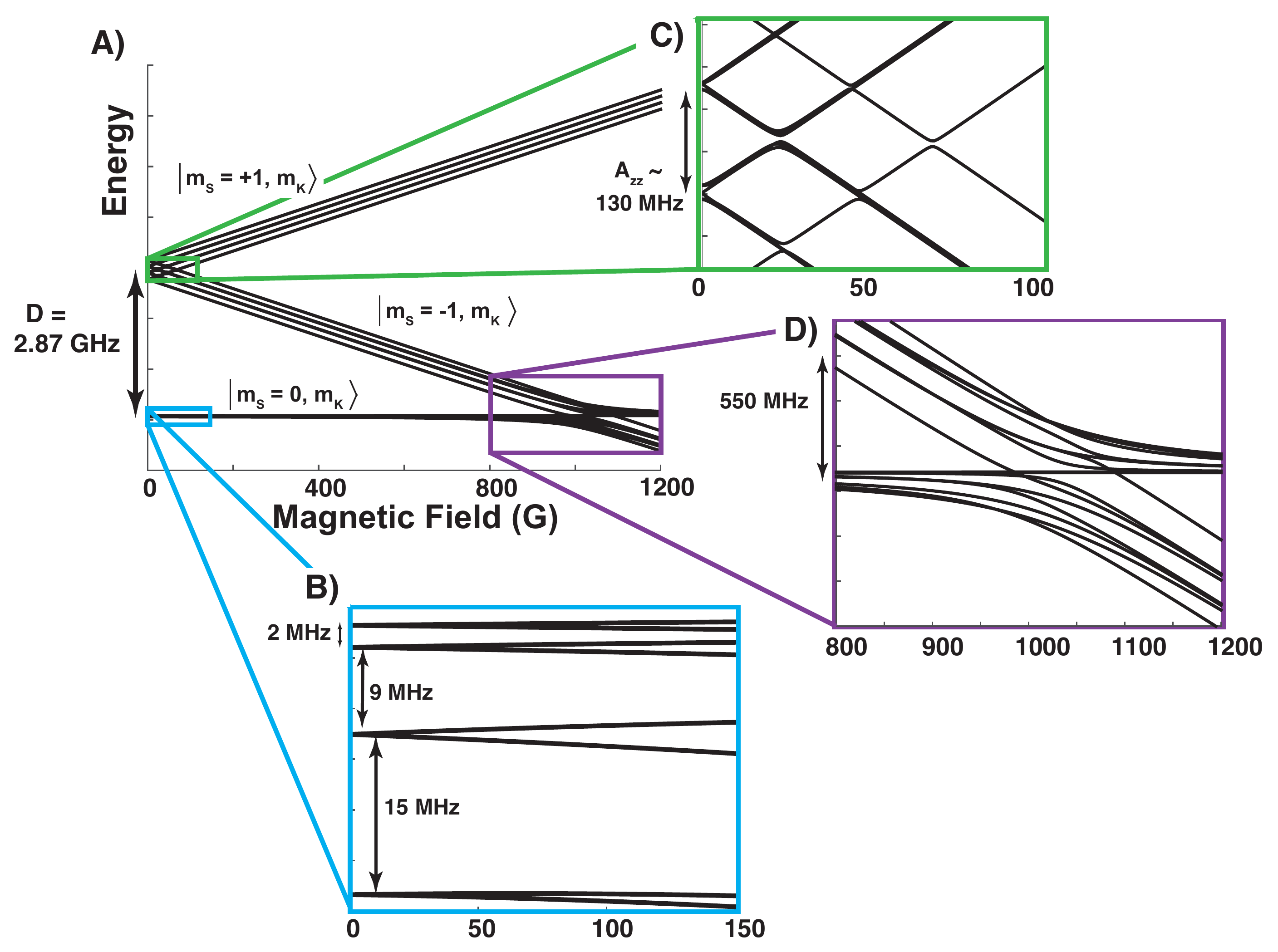}
\caption{(A) Calculated eigenvalues of the NV$^-$-(3)$^{13}$C spin system as a function of magnetic field strength along the NV$^-$ axis. (B) A more detailed view of the m$_S$ = 0 manifold is shown. The transverse terms of the hyperfine tensor result in four doubly-degenerate sublevels at zero magnetic field. This degeneracy is broken upon application of a magnetic field, and the eight sublevels exhibit a very weak, nonlinear magnetic field dependence from approximately 0-800 G. Two sets of LACs exist for the electronic ground state spin Hamiltonian of the NV-(3)$^{13}$C system. (C) Set 1 (0.5 - 80 G) occurs between the m$_S$ = -1 and m$_S$ = +1 manifolds and (D) Set 2 (800 - 1200G) occurs between the m$_S$ = -1 and m$_S$ = 0 manifolds. }
\label{fig:elevels}
\end{figure}

Two main sets of LACs where DPTs may be found are observed upon inspection of the eigenvalues of the NV$^-$-(3)$^{13}$C system in Figure \ref{fig:elevels}. Set 1 of LACs (Figure \ref{fig:elevels}C) results from the Zeeman interaction canceling the hyperfine interactions with the first-shell $^{13}$C nuclei, and occurs at magnetic field strengths 0-$\sim$80 G. Set 2 (Figure \ref{fig:elevels}D) results from the Zeeman interaction canceling the energy of the spin-spin coupling that leads to the ZFS. While a third set of LACs is known to occur from the spin states of the electronic excited state near 500 G, the hyperfine tensors of the NV$^-$ center and neighboring nuclear spins in the electronic excited state are largely unknown. In this study, we develop a simple basis for understanding the DPTs that become observable by CW-ODMR at magnetic fields approaching but not directly coinciding with the conditions of Set 2 of LACs ($\sim$565 - 950 G), due to the complexity of the behavior of eigenstates in that magnetic field regime (see the inset of Figure \ref{fig:predtrans}). The majority of these transitions have energies corresponding to transitions between eigenstates in the $m_S$ = 0 manifolds.

It is important to distinguish the transitions we are able to observe in this system using CW-ODMR from ZEFOZ transitions described thus far in the literature. In previous studies, ZEFOZ transitions are observed in spherically symmetric systems only perturbed by isotropic hyperfine interactions. For these systems, there exists a ``ZEFOZ point" at the LAC where the magnetic field dependence of at least one eigenstate involved in the transition approaches zero \cite{zhong15}. In contrast, our system is perturbed both by hyperfine interactions and the spin-spin coupling of the NV$^-$.  The transitions we study, which we distinguish by terming them decoherence-protected, occur between hyperfine sublevels of the m$_S$ = 0 manifold, whose magnetic field dependence originates only from mixing with m$_S$ = -1 eigenstates due to the transverse terms of the hyperfine interaction. This means that although a ZEFOZ point exists for these transitions at zero magnetic field where the coherence properties are optimal, there is a broad range of magnetic fields where transitions with enhanced coherence properties may be observed.  Furthermore, our detection method (CW-ODMR), though highly advantageous for its simplicity, is limited in the sense that it requires a transition to have a nonzero change in the z projection of electron spin angular momentum ($\Delta m_S = $) for observation. Transitions with the best coherence properties, where $\Delta m_S$ is closest to 0, are thus not directly observable via optical contrast. 

\begin{figure}[h]
	 \includegraphics[width=\textwidth]{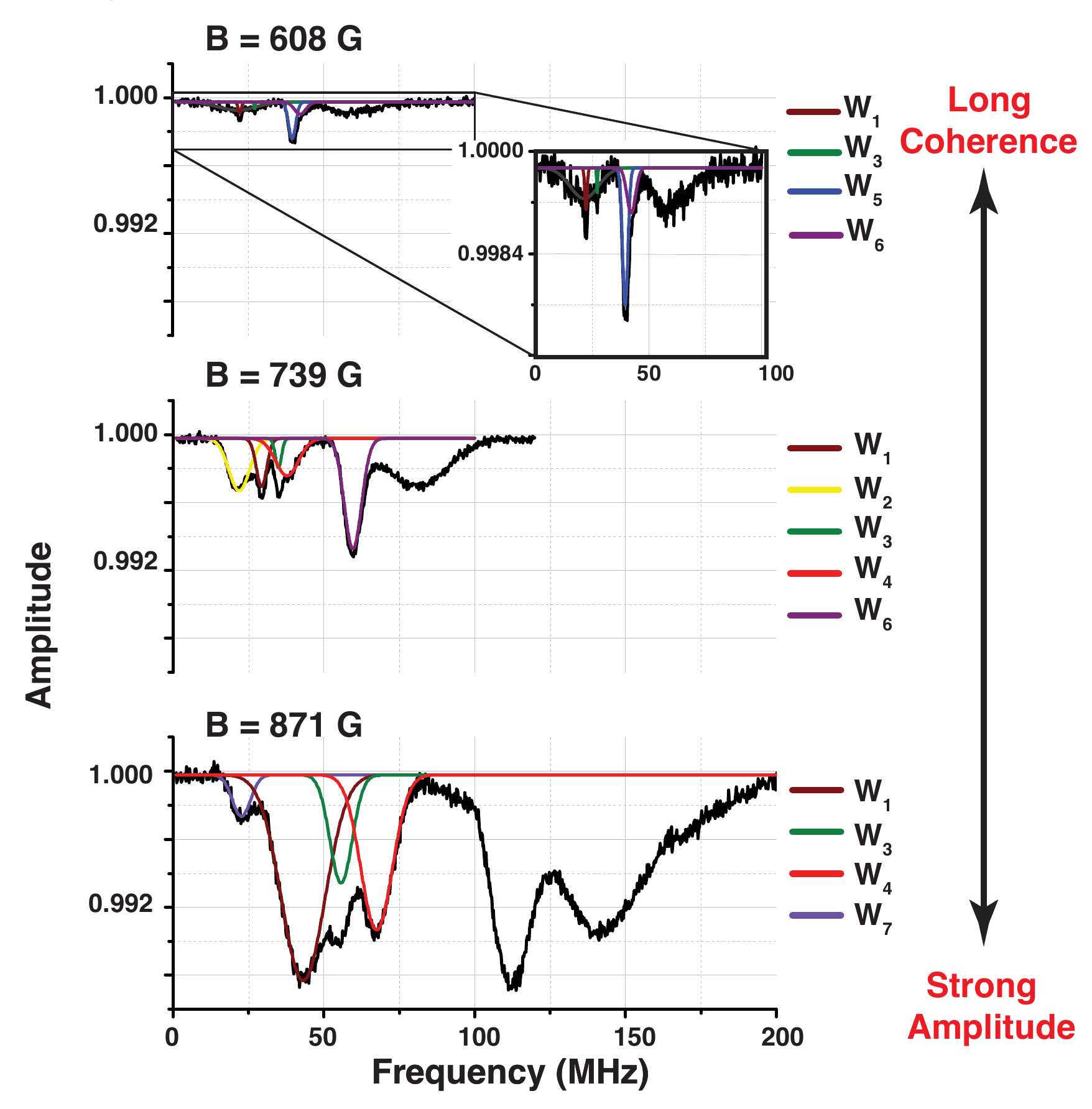}
	\caption{CW-ODMR spectra of DPTs for three different strengths of magnetic field are given to demonstrate how the nature of the transitions changes as the magnetic field is increased towards the regime of Set 2 of LACs. Gaussian fits of transitions occurring between the same sets of eigenstates (W$_i$) are color-coded. Unassigned transitions are not fit. Though the amplitude of transitions detectable by CW-ODMR far from Set 2 of LACs is low, the best coherence properties are exhibited in this magnetic field regime. }
	\label{fig:spectra}
\end{figure}

Nevertheless, we find DPTs between m$_S$ = 0 sublevels with sufficient optical contrast for CW-ODMR at magnetic fields as far from Set 2 of LACs as $\sim$550 G, where $\Delta m_S$ is on the order of 10$^{-3}$. The enhancement of coherence is gauged by the linewidths of these transitions, which are significantly narrowed in comparison to typical electron-type transitions of this spin system. The reduction in linewidth becomes less dramatic as the magnetic field approaches Set 2 of LACs, where the m$_S$ = 0 eigenstates become increasingly mixed with those in the m$_S$ = -1 manifold and transitions involve increasing $\Delta m_S$. This trend is reflected in the nonlinearity of transitions near Set 2 of LACs illustrated in the inset of Figure \ref{fig:predtrans}, as well as in Figure \ref{fig:spectra}, which shows spectra of DPTs selected over a range of magnetic fields (608-871G). It is clear that transitions with increasing optical contrast are also more sensitive to magnetic noise.

\begin{figure}[h]
	 \includegraphics[width=\textwidth]{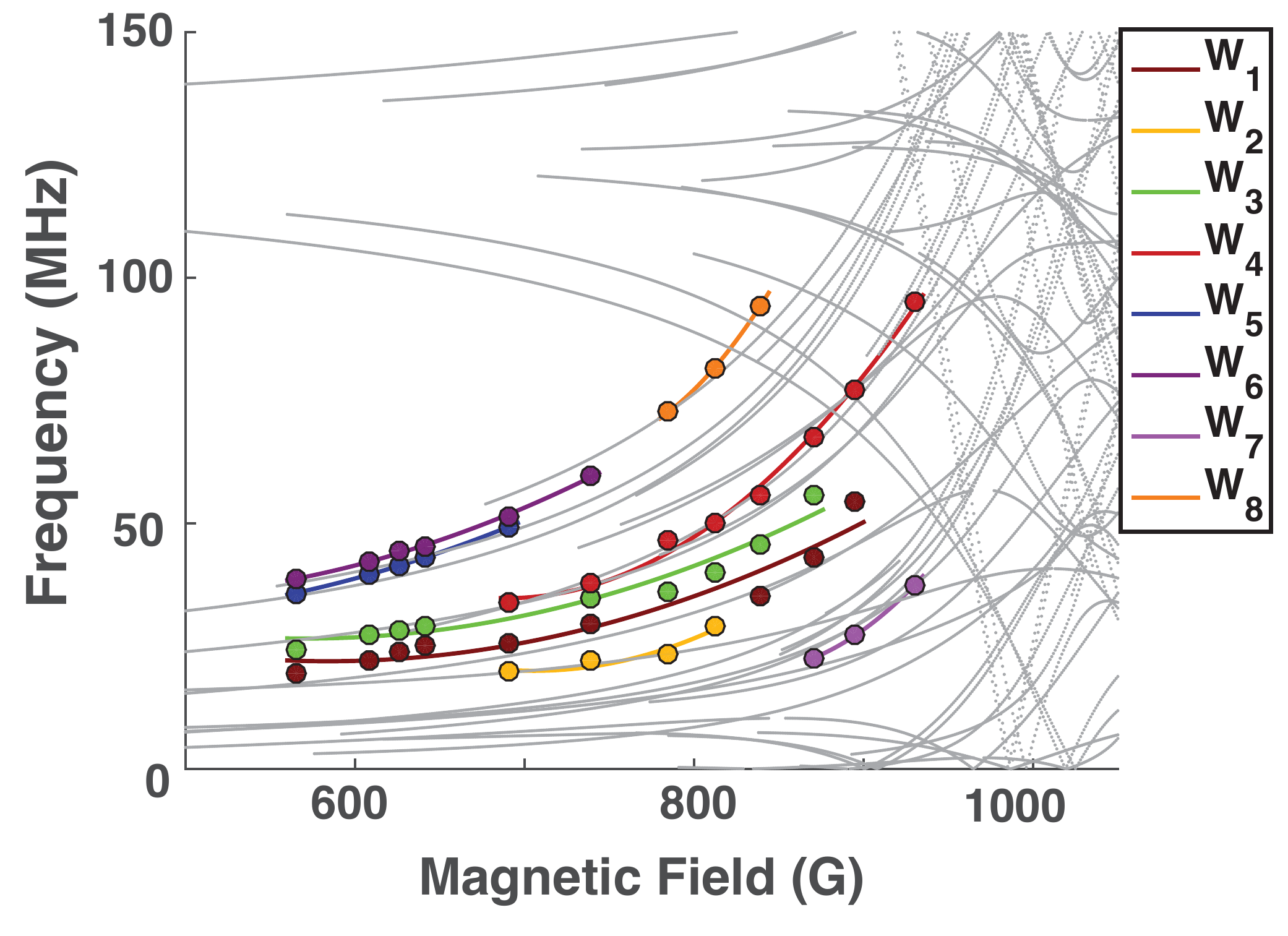}
	\caption{Assigned DPTs (W$_i$) of the NV$^{-}$-(3)\textsuperscript{13}C spin system compared to all predicted transitions at magnetic field strengths near Set 2 of LACs. The observed transitions corresponding to the same sets of eigenstates, as well as quadratic fits of their magnetic field dependence, are color-coded consistently with those in Figure \ref{fig:spectra}. Quadratic fits are used for empirical estimation of first- and second-order magnetic field dependence. The predicted transitions corresponding to $\kappa > 10^{-6}$ are underlaid in grayscale.}
	\label{fig:efgam}
\end{figure}

The observed transitions are assigned according to energy and the intensity factor $\kappa$ of the predicted transition (see Figure \ref{fig:efgam}). Transitions occurring between the same two eigenstates were grouped according to these assignments, and their first- $\left(\gamma_{eff} = \frac{\partial\nu}{\partial B} \right)$ and second-order $\left(curvature, C = \frac{\partial^2\nu}{\partial^2 B} \right)$ magnetic field dependence were determined empirically using quadratic fits. These parameters for the transitions shown in Figure \ref{fig:spectra} are given in Table \ref{tab:efgam}, and may be used to gauge the transition's sensitivity to axial magnetic field fluctuations. In this class of transitions, linewidths as narrow as 527 kHz as well as effective gyromagnetic ratios and curvatures as low as 13.62 kHz/G and 0.51 kHz/G$^2$ are estimated, respectively. Comparing with the maximum linewidth of 70.46 MHz for an electron-type transition we observe between the $m_S$ = 0 and $m_S$ = +1 eigenstates away from any set of LACs, these transitions are narrowed up to a factor of 133.7. The dephasing times $T_2^*$ estimated from these linewidths may be extended by use of dynamical decoupling methods \cite{dobrov13}.
 
Upon inspection of Figures \ref{fig:predtrans} and \ref{fig:efgam}, it is clear that the system is too complex to unambiguously assign each transition. Furthermore, direct comparisons between the observed linewidths and the effective gyromagnetic ratios characterizing a transition are not always consistent. This may indicate homogeneous broadening, which is not taken into account in the four-spin model. Inherent in this simplification of the spin system under consideration is that all sources of noise causing spectral broadening are axial magnetic fields that are static on the time scale of the NV$^-$-(3)$^{13}$C spin dynamics. In reality, the NV$^-$ experiences coupling to nuclei occupying all surrounding lattice sites in a volume with a radius as great as 5 angstroms from the defect.  Experimentally, 15 nuclear sites with interaction strengths ranging from 400 kHz to 14 MHz have been measured, whereas 9 nuclear sites with a total of 39 symmetrically equivalent positions have been calculated to have interaction strengths ranging from 1.5-19.4 MHz \cite{smeltz11,gali08}. Such a range of coupling provides various relaxation pathways induced by coherent interactions, i.e. nuclear spin flips, or energy conserving flip-flops, the understanding of which is beyond the scope of this manuscript.


Here the general behavior of a class of transitions in a novel spin system has been predicted. Whereas the amplitude of the transition observed by CW-ODMR becomes stronger near Set 2 of LACs, the coherence properties of DPTs approach an optimum far from Set 2 of LACs, where first- and second-order magnetic field dependence approaches a minimum. These trends are only true, however, for the DPTs we observe using CW-ODMR spectroscopy due to Set 2 of LACs. As is evident from Figures \ref{fig:elevels} and \ref{fig:predtrans}, this spin system is sufficiently complex to yield DPTs at a variety of magnetic field strengths and orientations. Narrow features at low magnetic fields (0.5-600 G, some examples found in the Earth's field CW-ODMR spectrum of NV$^-$ centers in 99\% $^{13}$C diamond in Figure \ref{fig:nvc13}) are discussed further in a study by Jarmola et. al. \cite{jarmola14}. Given the nature of DPTs, where eigenstates are least sensitive to inhomogeneous broadening in the limit of $\langle S_z \rangle \rightarrow 0$, alternate detection schemes such as raman heterodyne spectroscopy \cite{wong83,longdell06} may be required to achieve the maximum possible extension of coherence time in this spin system.

\section{Conclusion}
We have shown that the complex behavior of nitrogen vacancy centers in $^{13}$C-enriched diamond may be approximated by considering the simple spin molecule of the NV$^-$ center coupled to the three nearest-neighbor $^{13}$C nuclei. The internal interactions of this spin system, i.e. the spin-spin coupling of the NV$^-$ center and hyperfine interactions with three neighboring nuclei result in two regimes of magnetic field where spin states undergo avoided crossings and DPTs may be found. Such transitions involve electron-nuclear mixed spin states where sufficient projection of NV$^-$ spin onto the mixed state as well as optical pumping to the m$_S$ = 0 manifold allow for their observation by CW-ODMR spectroscopy. In this study we find a subset of DPTs whose linewidths are narrowed as a consequence of transitions in spin state being desensitized to the fluctuating magnetic field environment of $^{13}$C-enriched diamond. Of these transitions, we estimate effective gyromagnetic ratios as low as 13.62 kHz/G, with linewidths narrowed by factors as high as 133.7. These results demonstrate a method by which NV$^-$ coherence times may be significantly extended, which is a necessary step in the development of NV$^-$ centers for use in quantum information processing. 
%

\section{Experimental}
All spectra were acquired using simple continuous-wave application of laser and microwave excitation to detect changes in steady-state populations using a homebuilt confocal microscopy system. Magnetic fields as high as 1175 G are applied in these measurements by mounting a neodymium permanent magnet in a goniometer for control of field orientation. The relevant microwave frequencies are applied using waveform generators. The waveforms are chopped using a microwave switch and amplified before being sent to a broadband microwave loop resting directly beneath the diamond sample. Optical excitation is performed with a 200mW pumped solid state 532 nm laser. The beam is switched by driving an acousto-optic modulator and focused with a 0.7 NA microscope objective to an approximate waist of 1.4 ${\mu}$m (confocal length $\sim$23.6 ${\mu}$m, excitation volume $\sim$40 ${\mu}$m$^{3}$). After necessary filtering and loss due to optics, the power applied to this volume is 4 kW/cm$^{2}$. The fluorescence that results is separated from the excitation beam using a dichroic mirror, long-pass filtered, and fiber-coupled to a single photon counting module for detection. To describe the acquisition of a CW-ODMR spectrum, we refer you to the inset from Figure \ref{fig:nvc13}B. The time period for microwave excitation and fluorescence acquisition is 1 ms. The fluorescence acquired during microwave excitation is normalized by fluorescence acquired after the NV$^-$ ensemble is reinitialized by optical pumping. The sequence for each microwave frequency in the spectrum is averaged 2000 times. 

The diamond sample used in these measurements was an electronic grade wafer of dimensions 1 x 2 x 0.5 mm grown by Element Six. The sample was grown by chemical vapor deposition to a thickness of 500 ${\mu}$m, where 30 ${\mu}$m of the thickness on one face was grown from 99\% $^{13}$C-enriched methane. The estimated concentration of nitrogen impurity in the enriched layer is $<$ 1 ppm with less than 10 ppb NV$^-$ centers distributed unevenly across the sample. These naturally-formed NV$^-$ centers in the $^{13}$C-enriched layer were studied in the present experiments. For determination of magnetic field alignment along the NV$^-$ axis, an ensemble of nitrogen vacancy centers was created in the opposite face of the diamond where the abundance of $^{13}$C nuclei is 1.1\%. The creation of this ensemble of NV$^-$ centers was accomplished by an implant by Innovion Corporation of 35 keV and 60 keV $^{14}$N ions to a fluence of 10$^{13}$ ions/cm$^{2}$, followed by an anneal under nitrogen atmosphere for 2 hours at 800$^{\circ}$ C. The sample is situated over the broadband microwave loop and the focal point of laser beam leaving the microscope objective. A stepper motor with 1 ${\mu}$m stepping resolution in X, Y, and Z dimensions is used to position the implanted ensemble to align the field at the focal point of the beam, and then stepped to position the $^{13}$C-enriched layer at the focal point for studying DPTs. The quality of field alignment was confirmed by a difference in displacement of the m$_S$ = 0 to -1 and +1 transitions from the ground state zero field splitting of no greater than 5 MHz, allowing us to estimate an error of $\pm$ 2 G. The data collected at 785 G is an exception, where the difference in displacement was $\sim$10 MHz. In addition, we estimate the error associated with using this implanted ensemble of NV$^-$ centers as a magnetometer by the following equation \cite{acostathesis}:
 \begin{equation*}
\delta B  \simeq \frac{1}{\gamma _{NV}}\frac{1}{\sqrt{N_St_mT_2^*}}
\end{equation*}
Where $\gamma_{NV}$ is the NV$^-$ gyromagnetic ratio, $N_S$ is the number of NV$^-$ spins, $t_m$ is the measurement time, and $T_2^*$ is the dephasing time. For the concentration and volume of excitation given above, $N_S \simeq 6.5$ x $10^4$. When approximating $t_m \simeq T_2^* \simeq (\Delta\nu)^{-1}$, where $(\Delta\nu)^{-1}$ is the linewidth of the ensemble given above, we obtain a $\delta B$ of 0.05G, which is negligible in comparison to the field alignment error. 

\section{Acknowledgements}
This work was supported by the U.S. Department of Energy, Office of Science, Basic Energy Sciences under Contract No. DE-AC02-05CH11231. We would like to acknowledge Daniel Twitchen and Matthew Markham at Element 6 as well as Dmitry Budker for contribution of the sample used in this work.  Thermal annealing of the diamond sample was carried out at the Molecular Foundry at Lawrence Berkeley National Laboratory, which is supported by Office of Science, Office of Basic Energy Sciences, of the U.S. Department of Energy under Contract No. DE-AC02-05CH11231. We are grateful to Tevye Kuykendall for help in performing this task.

\bibliography{NV_Decoherence_Protected}


\section{All figures and tables}
 
 \begin{table}[h]\footnotesize
 \caption{\label{tab:efgam} Parameters of assigned decoherence protected transitions at 608 G, 739 G, and 871 G. Center frequencies ($\nu$) and linewidths ($\Delta\nu$) are given in MHz. $\gamma_{eff}$ $\left(\frac{\partial\nu}{\partial B}\right)$ and $C$ $\left(\frac{\partial^2\nu}{\partial^2B}\right)$ are given in kHz/G and kHz/G$^2$, respectively. For comparison, $\gamma_{NV}$ $\sim$2,800 kHz/G. Observed reductions in linewidth ($\epsilon$) are calculated using the average $\Delta\nu$ of electron-type transitions observed at magnetic fields away from any set of LACs (65.19 MHz).}
  \begin{ruledtabular}
 \begin{tabular}{c | c | c | c | c | c | c}
$B$ & $\nu$ & $\Delta\nu$ & ${\gamma}_{eff}$& $C$ & $\kappa$ & $\epsilon$ \\ \hline
\multirow{4}{*}{608} & 22.30 & 1.24 & 13.62 &  0.57 & 6.04 x 10$^{-6}$ & 52.52  \\
& 27.25 & 0.53 & 21.05 &  0.56 & 7.54 x 10$^{-5}$ & 123.69   \\
& 39.65 & 2.80 & 97.13 &  0.51 & 1.24 x 10$^{-4}$ & 23.32  \\
& 42.32 & 4.49 & 92.98 & 0.59 & 2.66 x 10$^{-6}$ & 14.51  \\ \hline
\multirow{5}{*}{739} & 21.89 & 7.88 & 49.45 & 1.48 & 1.38 x 10$^{-4}$ & 8.26 \\
& 29.35 & 4.26 & 87.90  & 0.57 & 3.11 x 10$^{-5}$ & 15.29 \\
& 34.93 & 2.46 & 94.65  & 0.56 & 2.98 x 10$^{-7}$ & 26.54 \\
& 37.93 & 8.46 & 101.28 & 2.05 & 3.18 x 10$^{-4}$  & 7.70 \\
& 59.65 & 6.63 & 170.30  & 0.59 & 4.94 x 10$^{-6}$ & 9.83 \\ \hline
\multirow{4}{*}{871} & 22.68 & 13.02 & 144.63 & 3.58 & 1.29 x 10$^{-3}$ & 9.14  \\
& 43.15 & 16.95 & 162.74 & 0.57 & 2.11 x 10$^{-4}$ & 3.85 \\
& 55.69 & 8.68 & 168.82 & 0.56 & 1.50 x 10$^{-3}$ & 7.51 \\
& 67.49 & 12.28 & 371.51 & 2.05 & 2.11 x 10$^{-5}$ & 5.31  \\
 \end{tabular}
 \end{ruledtabular}
 \end{table}

\end{document}